\documentclass[12pt]{article}

\usepackage[pdftex]{graphicx}
\usepackage{cite}

\usepackage{multirow}%
\usepackage{amsmath,amssymb,amsfonts}%
\usepackage{amsthm}%
\usepackage{mathrsfs}%
\usepackage{xcolor}%
\usepackage{authblk}
\usepackage{hyperref}
\usepackage{microtype}

\usepackage{hyperref}
\usepackage{lineno}
\usepackage{slashed} 
\usepackage{Spinor} 
\usepackage{DiSpinor} 
\usepackage{ReOperators}
\usepackage{Clifford}
\usepackage{StandardModel}
\usepackage{ExDelim}
 
\begin{document}

\title{Hard bremsstrahlung for the production of prompt photons}

\author[,1,2]{A.\,Kampf \footnote{E-mail: kampf@jinr.ru}}

\affil[1]{\small Dzhelepov Laboratory of Nuclear Problems, Joint Institute for Nuclear Research, Dubna, 141980, Russia}
\affil[2]{\small Lomonosov Moscow State University, Moscow, 119191, Russia}
 

\maketitle

\abstract{
The analytical calculation of helicity amplitudes for hard photon and gluon bremsstrahlung radiation are described using the spinor formalism for processes of the type $\bar{q} q \to g \gamma$ and $q g \to q \gamma$. The presented amplitudes are expressed in terms of field strength bivectors, with allowance for the dependence on the helicities and masses of all particles.
Numerical calculations have been performed for the quark-antiquark annihilation process $\bar{c}c$ at the center-of-mass energies of 24, 91, and 500 GeV, considering various combinations of helicities of the quarks and the photon in the final state.
This study is part of the ongoing development of theoretical support, based on the {\tt SANC} system, for the analysis of polarized observables in proton collisions at the \texttt{NICA} collider.
}
\section{Introduction}\label{intro}

The Nuclotron-based Ion Collider fAcility (NICA) \cite{SPD:2024gkq} is a collider that enables the study of polarized proton-proton collisions at the center-of-mass energies $\sqrt{s} = 24$–$27$ GeV. In this energy range, polarized gluon distribution functions remain poorly explored. The evaluation of polarization effects in proton-proton reactions should be carried out with the highest possible precision, since only a small fraction of partons are polarized, and as a result, the signal-to-background ratio is usually just a few percent. The most suitable processes for studying polarized gluon distributions are those involving prompt photon production. Prompt photons are produced through quark-antiquark annihilation and quark-gluon Compton scattering processes
\begin{gather}
\begin{aligned}
    \label{qqgA}
        &\bar{q}(p_1,\chi_1)q(p_2,\chi_2) \to 
        g(p_3,\chi_3)\gamma(p_4,\chi_4)(\gamma/g(p_5,\chi_5)),
        \\
        &q(p_1,\chi_1)g(p_2,\chi_2) \to  q(p_3,\chi_3)\gamma(p_4,\chi_4)(\gamma/g(p_5,\chi_5)),
    \end{aligned}
\end{gather}
where $q(\bar{q})$ is a quark (antiquark), $\gamma$ is a photon, $g$ is a gluon, and $p_{i}$ and $\chi_{i}$ ($i$ = 1 ... 5) represent the momentum and helicity, respectively.
Prompt photons carry information directly about the hard scattering process, which makes them especially valuable for analysis. The corresponding events are characterized by high purity, since the energy resolution of electromagnetic calorimeters is typically superior to that of hadronic calorimeters, and, consequently, systematic uncertainties are smaller.

The theoretical description of prompt photon production in hadronic collisions for the unpolarized case has been carried out in many studies. For example, radiative corrections at next-to-leading order in perturbative QCD were described in \cite{Aurenche:1987fs, Baer:1990ra, Gordon:1994ut}. The works \cite{Owens:1986mp, Berger:1990et} are the most comprehensive and contain extensive bibliographies. They include comparisons of theoretical predictions with experimental data, as well as a critical assessment of various sources of theoretical uncertainty. The polarized case is most thoroughly described in \cite{Gordon:1993qc}.

The estimation of observables at the leading order in perturbation theory for longitudinally polarized proton-proton collisions at \texttt{RHIC} \cite{Bunce:2000uv, Harrison:2003sb, RHICSPIN:2023zxx} was performed using the Monte Carlo generator \texttt{Sphinx} \cite{Guellenstern:1993qa, Gullenstern:1994sf, Gullenstern:1996pw}. Gluon distributions in the proton were fitted both at {\tt RHIC} \cite{Bourrely:1990pz} and at {\tt HERA} \cite{Aurenche:1989gv}.

Previously, calculations of polarized observables at the Born level \cite{Guskov:2019lvz, Saleev:2022jda, Saleev:2023shl, Alizada:2024nsr} were already performed for prompt photon production processes at \texttt{NICA}.

To date, the Monte Carlo (MC) generator {\tt ReneSANCe} \cite{Sadykov:2020any, Bondarenko:2022mbi, Sadykov:2023azk} enables the analysis of polarized observables for processes (\ref{qqgA}) at the Born level, taking into account longitudinal polarization. Our first step was the analysis of polarized observables and a comprehensive cross-check with existing third-party codes. The results were presented at the SPD collaboration workshop and the AYSS-2024 conference \cite{SPD:20241, SPD:20242}. Full agreement was achieved with the \texttt{Sphinx} program, while a discrepancy was found with the results of \cite{Saleev:2022jda, Saleev:2023shl}. Errors were identified in those works, which the authors corrected after a thorough comparison with our results.

Radiative corrections involve the calculation of one-loop amplitudes, their renormalization, the regularization of infrared divergences, and the combination of contributions from virtual and real emission. The complete one-loop cross section of the process can be divided into four components: the Born cross section, the contribution from virtual (loop) corrections, and the contributions from real corrections corresponding to soft and hard photon (or gluon) emission.

This work presents a description of the analytical calculations of the contribution from hard gluon (photon) bremsstrahlung and a validation of the result obtained for the annihilation process $\bar{c}c \to g\gamma g$.

In {\tt SANC}, each contribution is described using the helicity amplitude method, which is extremely convenient for taking into account the polarization of the initial and final states. To obtain covariant expressions for the helicity amplitudes of the virtual contribution, the internal procedure of the {\tt SANC} system \cite{Andonov:2004hi} based on the Vega-Wudka approach \cite{Vega:1995cc} is used. However, for the contribution from hard bremsstrahlung, the spinor formalism \cite{Kleiss:1985yh, Dittmaier:1998nn, Dittmaier:1993da} is employed.

The content of this work consists of three sections. Following the Introduction, Section 2 provides a detailed description of the analytical calculation of the hard bremsstrahlung amplitude in the spinor formalism for the process $\bar{q}qg\gamma(g)\to 0$. Numerical results are presented in Section 3. Section 4 is conclusion.

\section{Bremsstrahlung contribution}
\label{section1}

As mentioned above, to compute the covariant expressions for the helicity amplitudes of bremsstrahlung, this work employs the spinor formalism described in~\cite{Dydyshka:2023ffw}, which allows calculating matrix elements for massive particles. This technique also uses the symmetry properties of gauge theories as an additional means to simplify the matrix elements.

The covariant form is convenient for several reasons. First, the polarization vector of a gauge boson is not a covariant object, whereas the field strength bivector $\textbf{F}$ is. Therefore, it is expected that matrix elements will appear simpler when expressed in terms of the field strength bivectors. Second, this form is numerically more stable due to the absence of significant numerical cancellations.

In \texttt{SANC}, the studied process is traditionally calculated via annihilation into the vacuum
\begin{gather*}
    \begin{aligned}
        \sum\limits_{i=1}^{5} p_{i} = 0,
    \end{aligned}
\end{gather*}
and then the desired channel can be obtained using crossing symmetry. By applying the condition that the sum of the momenta is zero, processes (\ref{qqgA}) can be written as
\begin{gather}
    \begin{aligned}\label{zero}
       \bar{q}(p_{1},\chi_{1})+q(p_{2},\chi_{2})
       +g(p_{3},\chi_{3})+\gamma(p_{4},\chi_{4})
       +(\gamma/g)(p_{5},\chi_{5}) \to 0.
    \end{aligned}
\end{gather}
 In this way, we cover both cases of bremsstrahlung involving either a photon or a gluon.
 
\begin{figure}
  \begin{minipage}[t]{0.45\linewidth}
    \centering
    \includegraphics[width=\linewidth]{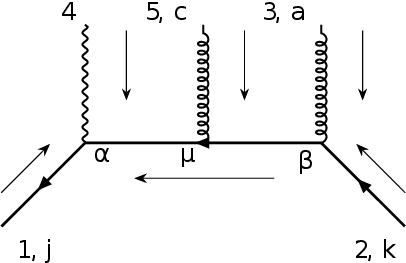} 
    \caption{One of the diagrams of process (\ref{zero}). The indices \textbf{c} and \textbf{a} correspond to different gluons, while \textbf{j} and \textbf{k} denote the color of the antiquark and quark. The vertex indices are \textbf{$\alpha$}, \textbf{$\beta$}, and \textbf{$\mu$}.}
    \label{fig:sum}
  \end{minipage}
  \hfill
\begin{minipage}[t]{0.45\linewidth}
    \centering
    \begin{minipage}[t]{\linewidth}
      \centering
      \includegraphics[width=\linewidth]{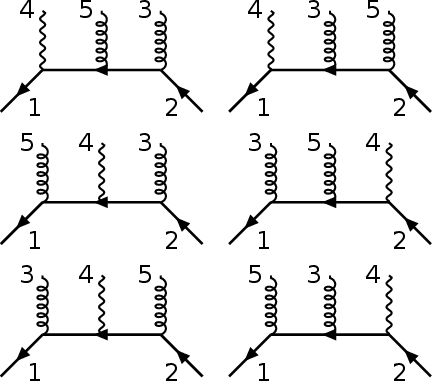}
      \caption{A subset of diagrams for the process with bremsstrahlung gluon emission, or the complete set of diagrams for process (\ref{zero}) with bremsstrahlung photon emission $g(p_{5},\chi_{5})\to \gamma(p_{5},\chi_{5})$.}
    \label{fig:base}
    \end{minipage}
    
    \vspace{1em} 
    
    \begin{minipage}[t]{\linewidth}
      \centering
      \includegraphics[width=\linewidth]{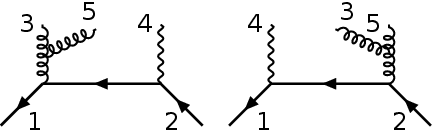}
      \caption{Diagrams with a three-gluon vertex for process (\ref{zero}) with bremsstrahlung gluon emission.}
    \label{fig:3gV}
    \end{minipage}
  \end{minipage}
\end{figure}

Let us consider the case of process (\ref{zero}) with bremsstrahlung gluon emission. Figure~\ref{fig:sum} shows one of the diagrams for this process, where arrows indicate the direction of momenta, indices \textbf{a} and \textbf{c} correspond to different gluons, while indices \textbf{k} and \textbf{j} denote the color of the quark and antiquark.

The complete set of diagrams for process (\ref{zero}) is presented in Figures~\ref{fig:base} and \ref{fig:3gV}. In the case where a bremsstrahlung photon is emitted instead of a gluon, only the set of diagrams shown in  Figure~\ref{fig:base} is required, with the gluon $g(p_{5},\chi_{5})$ replaced by the photon $\gamma(p_{5},\chi_{5})$, since the three-photon vertex is absent in the Standard Model. The amplitude for bremsstrahlung photon emission $\mathcal{A}^{\gamma}$ (\ref{Agamma}) consists of six terms corresponding to six permutations of two photons and one gluon.

In turn, the amplitude for bremsstrahlung gluon emission $\mathcal{A}^{g}$ (\ref{Ag}) consists of two contributions, $\mathcal{A}^{g}_{ac}$ and $\mathcal{A}^{g}_{ca}$, in which
\begin{gather*}
    \begin{aligned}
\gsq p_{ij...} = \slashed p_{i}+ m_{i}+\slashed p_{j}+ m_{j} + ...,\quad
\bgsq p_{ij...} = \slashed p_{i}-m_{i}+\slashed p_{j}-m_{j}+...,
    \end{aligned}
\end{gather*}
where $e$ is the positron charge, $Q$ is the charge of the quark (or antiquark), and $g_{s}$ is the strong coupling constant. To explicitly separate these contributions, the anti-commutation relation of the Gell-Mann matrices $[t^{a}, t^{c}] = i f^{acb} t^{b}$ was used.
\begin{gather}
    \begin{aligned}\label{Agamma}
    \mathcal{A}^{\gamma} = 
&
    (eQ)^2g_{s}it^{a}_{jk}\bar{v}_{1}\Bigg( \slashed\varepsilon_4 \dfrac{1}{\bgsq p_{235} }\slashed\varepsilon_5\dfrac{1}{\bgsq p_{23} } \slashed\varepsilon_3 
+
     \slashed\varepsilon_4 \dfrac{1}{\bgsq p_{235} } \slashed\varepsilon_3\dfrac{1}{\bgsq p_{25} } \slashed\varepsilon_5  
+
     \slashed\varepsilon_5 \dfrac{1}{\bgsq p_{234} } \slashed\varepsilon_4\dfrac{1}{\bgsq p_{23} } \slashed\varepsilon_3 
\\&  
+
     \slashed\varepsilon_3 \dfrac{1}{\bgsq p_{245}   }\slashed\varepsilon_5\dfrac{1}{\bgsq p_{24} } \slashed\varepsilon_4   
+   
      \slashed\varepsilon_3 \dfrac{1}{\bgsq p_{245} } \slashed\varepsilon_4\dfrac{1}{\bgsq p_{25} } \slashed\varepsilon_5 
+ 
     \slashed\varepsilon_5 \dfrac{1}{\bgsq p_{234} } \slashed\varepsilon_3\dfrac{1}{\bgsq p_{24} } \slashed\varepsilon_4
\Bigg)u_{2}.
    \end{aligned}
\end{gather}
\begin{gather}
\begin{aligned}\label{Ag}
\mathcal{A}^{g} = (eQ)g_{s}^{2}(\mathcal{A}_{ac} + \mathcal{A}_{ca}),
\\
\mathcal{A}_{ac} = 
-i(t^{a}t^{c})_{jk}\bar{v}_{1}\Bigg(
    \slashed\varepsilon_3 \dfrac{1}{\bgsq p_{245} } \slashed\varepsilon_4\dfrac{1}{\bgsq p_{25} } \slashed\varepsilon_5 
+
    \slashed\varepsilon_3 \dfrac{1}{\bgsq p_{245} }\slashed\varepsilon_5\dfrac{1}{\bgsq p_{24} } \slashed\varepsilon_4   
+ 
    \slashed\varepsilon_4 \dfrac{1}{\bgsq p_{235} } \slashed\varepsilon_3\dfrac{1}{\bgsq p_{25} } \slashed\varepsilon_5 
\\
-
   \gamma_{\alpha}\dfrac{1}{p^{2}_{241}}\big[(p_{3}-p_{241})^{\nu}g^{\alpha\rho}+(p_{241}-p_{5})^{\rho}g^{\alpha\nu}+(p_{5}-p_{3})^{\alpha}g^{\nu\rho}\big]\varepsilon_{3,\rho}\varepsilon_{5,\nu}\dfrac{1}{\bgsq p_{24}}\slashed\varepsilon_{4}
\\
-
    \slashed\varepsilon_{4}\dfrac{1}{\gsq p_{14}}\gamma_{\beta}\dfrac{1}{p^{2}_{241}}\big[(p_{241}-p_{3})^{\nu}g^{\beta\rho}+(p_{5}-p_{241})^{\rho}g^{\beta\nu}+(p_{3}-p_{5})^{\beta}g^{\nu\rho}\big]\varepsilon_{3,\rho}\varepsilon_{5,\nu}
\Bigg)u_{2},
\\
    \mathcal{A}_{ca} = 
-i(t^{c}t^{a})_{jk}\bar{v}_{1}\Bigg(
    \slashed\varepsilon_5 \dfrac{1}{\bgsq p_{234} } \slashed\varepsilon_4\dfrac{1}{\bgsq p_{23} } \slashed\varepsilon_3 
+
    \slashed\varepsilon_5 \dfrac{1}{\bgsq p_{234} } \slashed\varepsilon_3\dfrac{1}{\bgsq p_{24} } \slashed\varepsilon_4 
+
    \slashed\varepsilon_4 \dfrac{1}{\bgsq p_{235} }\slashed\varepsilon_5\dfrac{1}{\bgsq p_{23} } \slashed\varepsilon_3 
\\ 
+
    \gamma_{\alpha}\dfrac{1}{p^{2}_{241}}\big[(p_{3}-p_{241})^{\nu}g^{\alpha\rho}+(p_{241}-p_{5})^{\rho}g^{\alpha\nu}+(p_{5}-p_{3})^{\alpha}g^{\nu\rho}\big]\varepsilon_{3,\rho}\varepsilon_{5,\nu}\dfrac{1}{\bgsq p_{24}}\slashed\varepsilon_{4}
\\
+
    \slashed\varepsilon_{4}\dfrac{1}{\gsq p_{14}}\gamma_{\beta}\dfrac{1}{p^{2}_{241}}\big[(p_{241}-p_{3})^{\nu}g^{\beta\rho}+(p_{5}-p_{241})^{\rho}g^{\beta\nu}+(p_{3}-p_{5})^{\beta}g^{\nu\rho}\big]\varepsilon_{3,\rho}\varepsilon_{5,\nu}
\Bigg)u_{2},
\end{aligned}
\end{gather}

To reduce the amplitudes $\mathcal{A}^{g}$ (\ref{Ag}) and $\mathcal{A}^{\gamma}$ (\ref{Agamma}) to a form containing field strength bivectors, it is necessary to regroup the terms so that only the momenta $p_{i,\mu}$ and the explicitly gauge-covariant antisymmetric tensors $F_{i,\mu\nu}$ remain. These tensors can naturally be expressed as elements of the Clifford algebra of Dirac matrices by contraction with $\gamma^{[\mu}\gamma^{\nu]} = \gamma^{\mu} \wedge \gamma^{\nu}$:
\begin{equation}
\label{FM}
\textbf{F}_{i}\equiv F_{i,\mu\nu}\gamma^{\mu}\gamma^{\nu} = \slashed p_{i} \wedge \slashed \varepsilon_{i}=\slashed p_{i}\slashed\varepsilon_{i}.
\end{equation}
It is easy to verify that the field strength bivectors $\textbf{F}_{i}$ remain unchanged in their structure under the transformation $\slashed\varepsilon \to \slashed\varepsilon + C \slashed p$.

In the case of QED, the tensor $F_{i,\mu\nu}$ is the electromagnetic field tensor. In our case, it corresponds to a part of the gluon field tensor $$F_{i,\mu\nu} = \partial_{\mu}t_{a}A^{a}_{i,\nu}-\partial_{\nu}t_{a}A^{a}_{i,\mu},$$
which is invariant only under the Abelian subgroup of gauge transformations. After switching to the momentum representation, the color dependence will be carried only by the matrices $t_{a}$, and the expression for the bivector $\textbf{F}_{i}$ will exactly coincide with (\ref{FM}). In turn, the remaining part of the gluon field tensor $$G_{i,\mu\nu} = ig_{s}[t_{b},t_{c}]A^{b}_{i,\mu}A^{c}_{i,\nu}$$ is expressed in terms of commutators of the bivectors $\textbf{F}_{i}$ when regrouping the terms of the amplitude.

To define the polarization vectors of gluons and photons, one can use the axial gauge. This does not contradict the fact that the amplitude itself was obtained in the Feynman gauge, since physical polarization states do not depend on the choice of the gauge. In the case under consideration, gluons and photons are massless, so there are only two physical states for each massless boson. With the axial gauge taken into account, these satisfy the conditions
$(p_{i}\varepsilon_{i}) = 0$ and $(g\varepsilon_{i}) = 0$,
where $g_{\mu}$ is the gauge vector.
Solving this system of equations for $\varepsilon_{i}$, together with the normalization condition $(\varepsilon_{i}\varepsilon_{j}^{\star}) = -\delta_{ij}$, we obtain the polarization vector in the axial gauge
\begin{align*}
 \slashed\varepsilon_{i} &=\dfrac{\grade{1}{ \slashed g \mathbf F_{i}} }{(g p_{i})} 
,\quad
\grade{1}{G} \equiv \Trbar[G\gamma^{\mu}]\gamma_{\mu},\quad
 \slashed\varepsilon_{i}(g_1)- \slashed\varepsilon_{i}(g_2) =
-\dfrac{\Trbar[\slashed g_1  \slashed g_2 \mathbf{F}_{i} ] }{( g_1 p_{i})( g_2  p_{i})}  \slashed p_{i}
,\quad
\Trbar = \dfrac{1}{4}\Tr
.
\end{align*} 
Changing the vector $g$ results in a gauge transformation.

As we will see below, the amplitudes will consist of terms proportional to $\textbf{F}_{i}$, $\textbf{F}_{i}\textbf{F}_{j}$, $\textbf{F}_{i}\textbf{F}_{j}\textbf{F}_{k}$, and a scalar part $\mathcal{S}$. To construct these terms, it is necessary to single out a pair of the momentum and polarization vectors in the amplitude, $\slashed p_{i}\slashed\varepsilon_{i} = \textbf{F}_{i}$.

As an example, let us consider the first term of the amplitude $\mathcal{A}^{g}_{ac}$. After a standard transformation, we obtain
\begin{gather*}
    \begin{aligned}
        \slashed\varepsilon_3 \dfrac{\gsq p_{245}}{p^{2}_{245} - m^{2} } \slashed\varepsilon_4\dfrac{\gsq p_{25}}{p^{2}_{25} - m^{2}} \slashed\varepsilon_5,
    \end{aligned}
\end{gather*}
where the required pairs of vectors $\gsq p_{245}\slashed\varepsilon_{4}$ and $\gsq p_{25}\slashed\varepsilon_{5}$ have been formed.
From these pairs, the corresponding bivectors $\textbf{F}_{i}$ will be extracted in the subsequent steps.
For $\slashed\varepsilon_{3}$, it is necessary to introduce an additional vector by inserting a unit term as follows:
\begin{gather*}
    \begin{aligned}
        \bar{v}_{1}
=
        \dfrac{\bar{v}_{1}(\gsq p_{1}\bgsq q + \gsq q \bgsq p_{1})}{2(\gsq p_{1}\bgsq q)} 
=
        -\dfrac{\bar{v}_{1} \bgsq p_{2} \gsq p_{2345}}{2(p_{1}p_{2})+2m^{2}},
    \end{aligned}
\end{gather*}
where 
\begin{gather*}
\begin{aligned}
    \gsq p_i &= \slashed p_i + m_i
    ,&
    \bgsq p_i &= \slashed p_i - m_i ,
  \end{aligned}\\
\begin{aligned}
    \gsq p_i \bgsq p_j + \gsq p_j \bgsq p_i = 2(\gsq p_i\bgsq p_j),\quad
     (\gsq p_i\bgsq p_j) &= (p_i p_j) - m_i m_j .
  \end{aligned}
\end{gather*} 
Such a transformation yields the pair $\slashed p_{1} = -\slashed p_{2345}$ with $\slashed\varepsilon_{3}$.
The momentum $\slashed q$ serves as a kind of "gauge" momentum. If we set it equal to $\slashed p_{2}$, the scalar part of the amplitude $\mathcal{S}$ vanishes due to the application of the Dirac equation:
$\bar{v}_{1}\mathcal{S}\bgsq p_{2}u_{2} = 0.$

Now, using the simplest algebraic operations, we will demonstrate how the amplitude $\mathcal{A}_{ac}$ (\ref{Ag}) is reduced to the desired form. By successively applying the anticommutation relation for gamma matrices and using the Dirac equation, we obtain the following sequence of transformations for the first term:
\begin{gather*}
\begin{aligned}
&\bar{v}_{1}\bgsq p_{2}
    [\slashed p_{2345}+m]\slashed\varepsilon_{3}[\slashed p_{245} + m]\slashed\varepsilon_4 [\slashed p_{25} + m]\slashed\varepsilon_5 u_2 
\\=& 
    \bar{v}_{1}\bgsq p_{2}
\bigg[
    [2(p_{245}\varepsilon_{3})+\textbf{F}_{3}]
    \big[\textbf{F}_{4}  + 2( p_{25}\varepsilon_4) + \slashed\varepsilon_4(-\slashed p_{25} + m) \big] 
    [\slashed p_{25} + m]\slashed\varepsilon_5 
\\&
-
z_{245}\slashed\varepsilon_{3}\slashed\varepsilon_{4}[\slashed p_{25} + m]\slashed\varepsilon_5
\bigg]    
    u_2
\\=& 
\bar{v}_{1}\bgsq p_{2}
\bigg[
    [2(p_{245}\varepsilon_{3})+\textbf{F}_{3}]\bigg[
    [2 (p_{25}\varepsilon_4) + \mathbf{F}_4 ][2(p_2\varepsilon_5) + \mathbf{F}_5 ] 
- 
    z_{25}\slashed\varepsilon_4 \slashed\varepsilon_5\bigg]
\\&
-
    z_{245}\slashed\varepsilon_{3}\slashed\varepsilon_{4}[2(p_2\varepsilon_5) + \mathbf{F}_5 ]
    \bigg]    
    u_2,
  \end{aligned}
\end{gather*} 
where
\begin{gather*}
    \begin{aligned}
  z_{i\dots j} &= \gsq p_{i\dots j}\cdot \bgsq p_{i\dots j}
  = p_{i\dots j}^2 - (m_i +\dots+ m_j)^2,\quad
  p_{i...j} &= p_{i} + ... + p_{j}.       
    \end{aligned}
\end{gather*}
After similar transformations are performed on the remaining terms, the intermediate amplitude takes the form:
\begin{gather}\label{Agm}
\begin{aligned}
\mathcal{A}_{ac}=&
    \dfrac{i(t^{a}t^{c})_{jk}\bar{v}_{1} \bgsq p_{2}}{2(p_{1}p_{2})+2m^{2}}\Bigg(
\\&
    \dfrac{[2(p_{245}\varepsilon_3)+\textbf{F}_{3}]}{{ z_{245} }}\Bigg[\dfrac{[2(p_{25}\varepsilon_4) + \textbf{F}_{4}][2(p_{2}\varepsilon_5) + \textbf{F}_{5}]}{ z_{25} }-\slashed\varepsilon_{4}\slashed\varepsilon_{5}\Bigg] 
    - 
    \dfrac{\slashed\varepsilon_{3}\slashed\varepsilon_{4} \gsq p_{25}\slashed\varepsilon_{5}}{z_{25}}
\\ 
+&
    \dfrac{[2(p_{245}\varepsilon_3)+\textbf{F}_{3}]}{{ z_{245} }}\Bigg[\dfrac{[2(p_{24}\varepsilon_5) + \textbf{F}_{5}][2(p_{2}\varepsilon_4) + \textbf{F}_{4}]}{ z_{24} }-\slashed\varepsilon_{5}\slashed\varepsilon_{4}\Bigg] -\dfrac{\slashed\varepsilon_{3}\slashed\varepsilon_{5} \gsq p_{24}\slashed\varepsilon_{4}}{z_{24}}
\\+& 
    \dfrac{[2(p_{235}\varepsilon_4)+\textbf{F}_{4}]}{{ z_{235} }}\Bigg[\dfrac{[2(p_{25}\varepsilon_3) + \textbf{F}_{3}][2(p_{2}\varepsilon_5) + \textbf{F}_{5}]}{ z_{25} }-\slashed\varepsilon_{3}\slashed\varepsilon_{5}\Bigg] - \dfrac{\slashed\varepsilon_{4}\slashed\varepsilon_{3} \gsq p_{25}\slashed\varepsilon_{5}}{z_{25}}
\\-&
   \gsq p_{2345}\dfrac{1}{p^{2}_{241}}\big[
   ((p_{3}-p_{241})\varepsilon_{5})\slashed\varepsilon_{3}
+
   ((p_{241}-p_{5})\varepsilon_{3})\slashed\varepsilon_{5}
+
    (\slashed p_{5}-\slashed 
    p_{3})(\varepsilon_{3}\varepsilon_{5})
    \big]
    \dfrac{\gsq p_{24}}{z_{24}}\slashed\varepsilon_{4}
\\-&
   \gsq p_{2345}\slashed\varepsilon_{4}\dfrac{\bgsq p_{14}}{z_{14}}\dfrac{1}{p^{2}_{241}}\big[
   ((p_{241}-p_{3})\varepsilon_{5})\slashed\varepsilon_{3}
+
    ((p_{5}-p_{241})\varepsilon_{3})\slashed\varepsilon_{5}
+   
    (\slashed p_{3}-\slashed p_{5})(\varepsilon_{3}\varepsilon_{5})
   \big]\Bigg)u_{2}.
\end{aligned}
\end{gather}
From the resulting intermediate form (\ref{Agm}), one can already see some of the terms $\textbf{F}_{i}\textbf{F}_{j}\textbf{F}_{k}$, $\textbf{F}_{i}\textbf{F}_{j}$, and $\textbf{F}_{i}$. The remaining terms are implicitly contained in the three-gluon vertex shown in Fig. \ref{fig:3gV}. In the process of bringing the amplitude to the desired form, similar differences of pairs of bivectors arise from the terms of the three-gluon vertex:
\begin{gather*}
    \begin{aligned}
&-
        2(p_{3}\varepsilon_{5}) \slashed p_{5}\slashed\varepsilon_{3}
+
        2(p_{5}\varepsilon_{3}) \slashed p_{3}\slashed\varepsilon_{5}
-
        2(p_{3}p_{5}) \slashed\varepsilon_{3}\slashed\varepsilon_{5}
+
         (\varepsilon_{3}\varepsilon_{5})[\slashed p_{5}\slashed p_{3}-\slashed p_{3}\slashed p_{5}] 
\\&        
         = \textbf{F}_{3}\textbf{F}_{5}-\textbf{F}_{5}\textbf{F}_{3} = [\textbf{F}_{3},\textbf{F}_{5}].
    \end{aligned}
\end{gather*}

Next, one should collect the terms with the same number of bivectors, $\textbf{F}_{i}\textbf{F}_{j}$ and $\textbf{F}_{i}$, and combine them into $\Trbar$ in such a way that it contains the remaining bivector from the three available ones.

As an example, we consider how to extract $\Trbar$ for the combination of the bivectors $\textbf{F}_{4}\textbf{F}_{5}$ and $\textbf{F}_{3}$. Collecting all the terms and taking into account that $z_{ijk}=z_{ij}-2(p_{ij}p_{k})$ and $p_{245} = - p_{13}$, we obtain the following sequence of transformations:
\begin{gather*}
    \begin{aligned}
    &\Bigg(
    \dfrac{2(p_{245}\varepsilon_{3})}{z_{245}z_{25}}
+ 
    \dfrac{2(p_{25}\varepsilon_{3})}{z_{235}z_{25}}
+
    \dfrac{2(p_{5}\varepsilon_{3})}{z_{14}z_{241}}
    \Bigg)
\textbf{F}_{4}\textbf{F}_{5}
=
\Bigg(
-
        \dfrac{2(p_{1}\varepsilon_{3})}{z_{25}z_{13}}
+
        \dfrac{2(p_{5}\varepsilon_{3})}{z_{25}z_{235}}
\\&
+
        \dfrac{2(p_{2}\varepsilon_{3})z_{35}+2(p_{5}\varepsilon_{3})[z_{235}-2(p_{25}p_{3})]}{z_{25}z_{35}z_{235}}
\Bigg)\textbf{F}_{4}\textbf{F}_{5}
=
\Bigg(
        \dfrac{2(p_{5}\varepsilon_{3})z_{13}-2(p_{1}\varepsilon_{3})z_{35}}{z_{13}z_{25}z_{35}}
\\&
+
        \dfrac{2(p_{2}\varepsilon_{3})z_{35}-2(p_{5}\varepsilon_{3})z_{23}}{z_{25}z_{35}z_{235}}
\Bigg)\textbf{F}_{4}\textbf{F}_{5}
=
\Bigg(
        \dfrac{\Trbar[\slashed p_{5}\slashed p_{1}\textbf{F}_{3}]}{z_{13}z_{25}z_{35}}
+
        \dfrac{\Trbar[\slashed p_{2}\slashed p_{5}\textbf{F}_{3}]}{z_{25}z_{35}z_{235}}
    \Bigg)
\textbf{F}_{4}\textbf{F}_{5}.
    \end{aligned}
\end{gather*}
For the terms with a single $\textbf{F}_{i}$, the process of obtaining $\Trbar$ remains the same. However, the number of terms increases, and so consequently does the complexity of selecting the right combination needed to extract $\Trbar$. If we take into account that $\Trbar[p\grade1{G}] = 4(p\grade1{G})$ (where $p$ is an arbitrary vector) and $p_{235} = -p_{14}$, then for $\textbf{F}_{3}$ we obtain
\begin{gather*}
    \begin{aligned}
&\Bigg(
        \dfrac{2(p_{25}\varepsilon_{4})2(p_{2}\varepsilon_{5})}{z_{245}z_{25}}
+
        \dfrac{2(p_{24}\varepsilon_{5})2(p_{2}\varepsilon_{4})}{z_{245}z_{24}}
+
        \dfrac{2(p_{235}\varepsilon_{4})2(p_{2}\varepsilon_{5})}{z_{235}z_{25}}
-
        \dfrac{2(\varepsilon_{4}\varepsilon_{5})}{z_{245}}
-
        \dfrac{2(p_{3}\varepsilon_{5})2(p_{2}\varepsilon_{4})}{z_{24}z_{241}}
\\&
-
        \dfrac{2(p_{235}\varepsilon_{4})2(p_{3}\varepsilon_{5})}{z_{14}z_{241}}
\Bigg)\textbf{F}_{3}
=
\Bigg(
        \dfrac{2(p_{2}\varepsilon_{5})}{z_{245}}\Bigg[\dfrac{2(p_{25}\varepsilon_{4})z_{24} + 2(p_{2}\varepsilon_{4})[z_{245}-2(p_{25}p_{4})]}{z_{24}z_{25}}
\Bigg]
\\&
+
        \dfrac{2(p_{4}\varepsilon_{5})2(p_{2}\varepsilon_{4})}{z_{245}z_{24}}
+
        \dfrac{2(p_{235}\varepsilon_{4})2(p_{2}\varepsilon_{5})}{z_{235}z_{25}}
-
        \dfrac{2(\varepsilon_{4}\varepsilon_{5})}{z_{245}}
-
        \dfrac{2(p_{3}\varepsilon_{5})2(p_{2}\varepsilon_{4})}{z_{24}z_{241}}
+
        \dfrac{2(p_{1}\varepsilon_{4})2(p_{3}\varepsilon_{5})}{z_{14}z_{241}}
\Bigg)\textbf{F}_{3}
\\
&=
\Bigg(
        \dfrac{2(p_{2}\varepsilon_{5})\Trbar[p_{25}p_{2}\textbf{F}_{4}]}{z_{24}z_{25}z_{245}}
+
        \dfrac{2(p_{2}\varepsilon_{5})}{z_{25}}
\Bigg[
        \dfrac{2(p_{2}\varepsilon_{4})z_{14}-2(p_{1}\varepsilon_{4})z_{24}}{z_{14}z_{24}}
\Bigg]
\\&
+
        \dfrac{2(p_{4}\varepsilon_{5})2(p_{2}\varepsilon_{4})
-
        2(\varepsilon_{4}\varepsilon_{5})z_{24}}{z_{24}z_{245}}
+
        \dfrac{2(p_{3}\varepsilon_{5})}{z_{241}}
        \Bigg[
        \dfrac{2(p_{1}\varepsilon_{4})z_{24}-2(p_{2}\varepsilon_{4})z_{14}}{z_{14}z_{24}}
        \Bigg]
\Bigg)\textbf{F}_{3}
\\
&=
\Bigg(
        \dfrac{2(p_{2}\varepsilon_{5})\Trbar[p_{25}p_{2}\textbf{F}_{4}]
-       
        z_{25}\Trbar[\varepsilon_{5}p_{2}\textbf{F}_{4}]}{z_{24}z_{25}z_{245}}
+
        \dfrac{2(p_{2}\varepsilon_{5})z_{35}\Trbar[p_{2}p_{1}\textbf{F}_{4}]}{z_{14}z_{24}z_{25}z_{35}}
\\&
-
        \dfrac{2(p_{3}\varepsilon_{5})z_{25}\Trbar[p_{2}p_{1}\textbf{F}_{4}]}{z_{14}z_{24}z_{25}z_{35}}
\Bigg)\textbf{F}_{3}
=
\Bigg(
        \dfrac{2\Trbar[p_{2}\grade1{p_{2}\textbf{F}_{4}}\textbf{F}_{5}]}{z_{24}z_{25}z_{245}}
+
        \dfrac{\Trbar[p_{2}p_{3}\textbf{F}_{5}]\Trbar[p_{2}p_{1}\textbf{F}_{4}]}{z_{14}z_{24}z_{25}z_{35}}
\Bigg)\textbf{F}_{3}.
\end{aligned}
\end{gather*}

By extracting all possible combinations of $\textbf{F}_{i}$ and collecting all terms with identical combinations of bivectors into $\Trbar$, we obtain the amplitude $\mathcal{A}^{g}_{ac}$ in the form:
\begin{gather}\label{Agf}
    \begin{aligned}
        \mathcal{A}_{ac} =&
        \dfrac{i(t^{a}t^{c})_{jk}\bar{v}_{1} \bgsq p_{2}}{2(p_{1}p_{2})+2m^{2}}
\Bigg[
        \dfrac{\textbf{F}_{3}\textbf{F}_{4}\textbf{F}_{5}}{z_{245}z_{25}}
+   
        \dfrac{\textbf{F}_{3}\textbf{F}_{5}\textbf{F}_{4}}{z_{245}z_{24}}
+
        \dfrac{\textbf{F}_{4}\textbf{F}_{3}\textbf{F}_{5}}{z_{235}z_{25}}
+
        \dfrac{\textbf{F}_{4}[\textbf{F}_{3},\textbf{F}_{5}]}{z_{14}z_{241}}
+
        \dfrac{[\textbf{F}_{3},\textbf{F}_{5}]\textbf{F}_{4}}{z_{24}z_{241}}
\\+&
    \Bigg(
        \dfrac{\Trbar[\slashed p_{2}\slashed p_{5}\textbf{F}_{3}]}{z_{25}z_{35}z_{235}}
+
        \dfrac{\Trbar[\slashed p_{5} \slashed p_{1}\textbf{F}_{3}]}{z_{13}z_{25}z_{35}}
    \Bigg)
\textbf{F}_{4}\textbf{F}_{5}
+
    \dfrac{\Trbar[\slashed p_{5}\slashed p_{1}\textbf{F}_{3}]}{z_{13}z_{24}z_{35}}\textbf{F}_{5}\textbf{F}_{4}
+
    \Bigg(
    \dfrac{\Trbar[\slashed p_{2}\slashed p_{3}\textbf{F}_{5}]}{z_{25}z_{241}z_{245}}
\\&
+
    \dfrac{\Trbar[\slashed p_{3}\slashed p_{1}\textbf{F}_{5}]}{z_{24}z_{241}z_{245}}
    \Bigg)
\textbf{F}_{3}\textbf{F}_{4}
+
    \dfrac{\Trbar[\slashed p_{2}\slashed p_{3}\textbf{F}_{5}]}{z_{14}z_{25}z_{35}}
\textbf{F}_{4}\textbf{F}_{3}
+
    \Bigg(
    \dfrac{\Trbar[\slashed p_{2}\slashed p_{1}\textbf{F}_{4}]}{z_{14}z_{24}z_{245}}
+
    \dfrac{\Trbar[\slashed p_{1}\slashed p_{3}\textbf{F}_{4}]}{z_{14}z_{25}z_{245}}
    \Bigg)
\textbf{F}_{3}\textbf{F}_{5}
\\+&
\dfrac{\Trbar[\slashed p_{2}\slashed p_{1}\textbf{F}_{4}]}{z_{14}z_{24}z_{241}}
[
        \textbf{F}_{3},\textbf{F}_{5}
]
+
\Bigg(
        \dfrac{2\Trbar[\slashed p_{2}\grade1{\slashed p_{2}\textbf{F}_{4}}\textbf{F}_{5}]}{z_{24}z_{25}z_{245}}
+
        \dfrac{\Trbar[\slashed p_{2}\slashed p_{1}\textbf{F}_{4}]\Trbar[\slashed p_{2}\slashed p_{3}\textbf{F}_{5}]}{z_{24}z_{25}z_{35}z_{235}}
\Bigg)\textbf{F}_{3}
\\+& 
\Bigg(
        \dfrac{2\Trbar[\slashed p_{2}\grade1{\slashed p_{1}\textbf{F}_{3}}\textbf{F}_{5}]}{z_{13}z_{35}z_{25}}
+
        \dfrac{2\Trbar[\grade1{\slashed p_{1}\textbf{F}_{3}}\slashed p_{1}\textbf{F}_{5}]}{z_{13}z_{35}z_{135}}
+
        \dfrac{2\Trbar[\grade1{\slashed p_{2}\textbf{F}_{3}}\slashed p_{2}\textbf{F}_{5}]}{z_{25}z_{35}z_{235}}
\Bigg)\textbf{F}_{4}
\\+&
\Bigg(
        \dfrac{\Trbar[\slashed p_{5}\slashed p_{1}\textbf{F}_{3}]\Trbar[\slashed p_{2}\slashed p_{1}\textbf{F}_{4}]}{z_{24}z_{25}z_{35}z_{245}}
+
        \dfrac{2\Trbar[\slashed p_{1}\grade1{\slashed p_{2}\textbf{F}_{4}}\textbf{F}_{3}]}{z_{24}z_{25}z_{245}}
+
        \dfrac{\Trbar[\slashed p_{2}\slashed p_{1}\textbf{F}_{4}]\Trbar[\slashed p_{2}\slashed p_{5}\textbf{F}_{3}]}{z_{24}z_{25}z_{35}z_{235}}
         \Bigg)\textbf{F}_{5}
\Bigg]u_{2},
    \end{aligned}
\end{gather}
here
\begin{gather*}
\begin{aligned}
  \grade1{\slashed p_{m}\textbf{F}_{j}} &= \dfrac{\slashed p_{m}\textbf{F}_{j}-\textbf{F}_{j}\slashed p_{m}}{2}.
  \end{aligned}
\end{gather*}
The amplitude with the opposite quark ordering, $\mathcal{A}_{ca}$, is obtained from the above one by swapping the indices $3 \leftrightarrow 5$.
The expression for the bremsstrahlung photon emission amplitude, $\mathcal{A}^{\gamma}$, is derived in a similar way and takes the following concise form:
\begin{gather}
        \begin{aligned}\label{Agammaf}
        &\mathcal{A}^{\gamma}_{a} = \dfrac{(eQ)^2g_{s}it^{a}_{jk}\bar{v}_{1} \bgsq p_{2}}{2(p_{1}p_{2})+2m^{2}}
\Bigg[
     \sum_{\substack{ i,j,k = \{3,4,5\}\\ i \neq j \neq k }}
\Bigg(
    \dfrac{\textbf{F}_{i}\textbf{F}_{j}\textbf{F}_{k}}{z_{2jk}z_{2k}}
+
    \textbf{F}_{i}\textbf{F}_{j}
\bigg[
    \dfrac{\textbf{Tr}[\slashed p_{2j}\slashed p_{2}\textbf{F}_{k}]}{z_{2k}z_{2j}z_{2jk}}+\dfrac{\textbf{Tr}[\slashed p_{2}\slashed p_{1}\textbf{F}_{k}]}{z_{2j}z_{2k}z_{1k}}
\bigg]
\Bigg)
\\&
+
    \sum_{\substack{i=3,j=4,k=5\\ i=4,j=5,k=3\\i=5,j=4,k=3}}
\textbf{F}_{i}\Bigg(
    \dfrac{\textbf{Tr}[\slashed p_{1}\slashed p_{2}\textbf{F}_{j}]\textbf{Tr}[\slashed p_{1}\slashed p_{2}\textbf{F}_{k}]}{z_{1j}z_{2j}z_{1k}z_{2k}}
+
    \dfrac{2\textbf{Tr}[\slashed p_{1}\grade1{\slashed p_{1}\textbf{F}_{j}}\textbf{F}_{k}]}{z_{1j}z_{1k}z_{1jk}}
+
    \dfrac{2\textbf{Tr}[\slashed p_{2}\grade1{\slashed p_{2}\textbf{F}_{j}}\textbf{F}_{k}]}{z_{2j}z_{2k}z_{2jk}}
\Bigg)
\Bigg]u_{2}.
    \end{aligned}
\end{gather}

It can be noticed that the amplitudes contain several basic components:
$\Trbar[\slashed p_{l}\textbf{F}_{i}\slashed p_{n}\textbf{F}_{j}]$, $\Trbar[\slashed p_{l}\slashed p_{m}\textbf{F}_{i}]$, $\Trbar[\textbf{F}_{i}\textbf{F}_{j}]$, $\textbf{F}_{i}\textbf{F}_{j}\textbf{F}_{k}$, $\textbf{F}_{i}\textbf{F}_{j}$ and $\textbf{F}_{i}$. To express them in terms of Dirac spinors, we will use the notation
\begin{gather*}
    \begin{aligned}
        \bar{v}_{i}\equiv\bar{v}^{\chi_{i}}(p_{i}) \equiv \SpVI{i^{\chi_{i}}}, \quad u_i \equiv u^{\chi_{i}}(p_{i}) \equiv \SpIU{i^{\chi_{i}}}, 
    \end{aligned}
\end{gather*}
and the fact that the bivector $\textbf{F}_{i}$ can be represented as $\textbf{F}^{\chi_{i}}_{i} = \sqrt{2}u^{\chi{i}}(p_{i})\bar{v}^{\chi_{i}}(p_{i})$,
then
\begin{gather}\label{Tr}
    \begin{aligned}
        \bar{v}_{q}\bgsq p_{2}\textbf{F}_{i}\textbf{F}_{j}u_{m} &= (\sqrt{2})^{2}\SpVI{q^{\chi_{q}}}\bgsq p_{2}\SpIU{i^{\chi_{i}}}\SpVIU{i^{\chi_{i}}}{j^{\chi_{j}}}\SpVIU{j^{\chi_{i}}}{m^{\chi_{m}}},
\\
        \bar{v}_{q}\bgsq p_{2}\Trbar[\textbf{F}_{i}\textbf{F}_{j}]u_{m} 
&=       
        (\sqrt{2})^{2}\SpVI{q^{\chi_{q}}}\bgsq p_{2}\SpIU{m^{\chi_{m}}} \SpVIU{j^{\chi_{j}}}{i^{\chi_{i}}}\SpVIU{i^{\chi_{i}}}{j^{\chi_{j}}},
\\
        \bar{v}_{q}\bgsq p_{2}\Trbar[p_{l}\textbf{F}_{i}p_{n}\textbf{F}_{j}]u_{m} 
&=
        (\sqrt{2})^{2}\SpVI{q^{\chi_{q}}}\bgsq p_{2}\SpIU{m^{\chi_{m}}}\SpVI{j^{\chi_{j}}}\slashed p_{l} \SpIU{i^{\chi_{i}}}\SpVI{i^{\chi_{i}}}\slashed p_{n} \SpIU{j^{\chi_{j}}},
\\
        \bar{v}_{q}\bgsq p_{2}\textbf{F}_{i}\textbf{F}_{j}\textbf{F}_{k}u_{m} 
&= 
        (\sqrt{2})^{3}\SpVI{q^{\chi_{q}}}\bgsq p_{2}\SpIU{i^{\chi_{i}}}\SpVIU{i^{\chi_{i}}}{j^{\chi_{j}}}\SpVIU{j^{\chi_{i}}}{k^{\chi_{k}}}\SpVIU{k^{\chi_{k}}}{m^{\chi_{m}}}
\\
        \bar{v}_{q}\bgsq p_{2}\Trbar[p_{l}p_{n}\textbf{F}_{i}]u_{m} 
&=   
        \sqrt{2}\SpVI{q^{\chi_{q}}}\bgsq p_{2}\SpIU{m^{\chi_{m}}}\SpVI{i^{\chi_{i}}}\slashed p_{l}\slashed p_{n}\SpIU{i^{\chi_{i}}}
    \end{aligned}
\end{gather}
Substituting expressions (\ref{Tr}) into amplitudes (\ref{Agf}) and (\ref{Agammaf}), we obtain the helicity amplitudes expressed in terms of Dirac spinors.

The subsequent calculations of the contributions from hard photon and gluon bremsstrahlung -- specifically, squaring the amplitudes and integrating over the phase space -- were performed numerically using the Monte Carlo tools {\tt SANC}: the generator \texttt{ReneSANCe} and the integrator \texttt{MCSANC} \cite{Bondarenko:2013nu}.

\section{Numerical results}
\label{section2}

As an example for verifying the analytical results, the amplitude $\mathcal{A}^{g}$ was considered in the annihilation channel (\ref{qqgA}) of a $\bar{c}c$ quark pair:
\begin{gather}
\begin{aligned}
    \label{ccA}
        &\bar{c}(p_1,\chi_1)c(p_2,\chi_2) \to 
        g(p_3,\chi_3)\gamma(p_4,\chi_4)(g(p_5,\chi_5)).
    \end{aligned}
\end{gather}
The calculations were performed using the
\begin{tabbing}
\hspace{3.5em}\=\hspace{1.8em}\=\hspace{9em}\=\hspace{1.8em}\=\hspace{1.8em}\=\hspace{1em}\kill
$\alpha^{-1}(0)$\> = \> 137.035999084,\> $m_c$\> = \> 1.67 GeV,\\
$\alpha_{s}(M_{Z})$  \>=\> 0.1178,\>  $M_Z$ \>=\> 91.1876 GeV,\\
$M_W$ \>=\> 80.379 GeV.
\label{parameters}
\end{tabbing}
and the kinematic cuts:
\begin{itemize}
    \item energy $g$ and $\gamma$ $>  0.0001\dfrac{\sqrt{s}}{2}$ GeV,
    \item $20^{\circ} < \theta_{g,\gamma} < 160^{\circ}$ for $g$ and $\gamma$,
    \item $M_{gg} > 1$ GeV,
\end{itemize}
where $s = (p_{1} + p_{2})^{2}$ is the Mandelstam variable, $\theta_{g,\gamma}$ is the scattering angle of the gluon or photon in the laboratory frame, and $M_{gg}$ is the invariant mass of the gluon pair. The $z$-axis is directed along the momentum of the antiquark.

The comparison was carried out with the external Monte Carlo codes \texttt{CalcHEP} \cite{Belyaev:2012qa} and \texttt{WHIZARD} \cite{Moretti:2001zz, Kilian:2007gr}.
The total cross section was calculated for the operating energies of \texttt{NICA} (24 GeV) and \texttt{RHIC} (500 GeV), and at the $Z$-boson resonance in the center-of-mass system for various helicity values of the $\bar{c}$ and $c$ quarks.
The results of the comparison are presented in Tables \ref{un}, \ref{hel24}, \ref{hel91}, and \ref{hel500}.
As seen in the tables, \texttt{ReneSANCe}, \texttt{CalcHEP}, and \texttt{WHIZARD} show good agreement in all cases considered, with consistency at the level of three to five significant digits within the margin of error.
With the inclusion of quark helicity it is seen that the $++$ and $--$ combinations are suppressed relative to the $+-$ and $-+$ combinations by one order of magnitude at 24 and 91 GeV, and by four orders of magnitude at 500 GeV.
\begin{table}[ht!]
\centering
\caption{
The total cross section of process (\ref{ccA}) in picobarns, averaged over helicities, at various center-of-mass energies.}
\label{un}
    \begin{tabular}{llll}
    \hline\hline
    $\sqrt{s}$, GeV & {24} & {91} & {500}\\
    \hline
    {\tt ReneSANCe}	          & $106(1)\cdot10^{2}$   & $133(1)\cdot 10$ & 69.8(1) \\  
    {\tt CalcHEP}             & $106(1)\cdot10^{2}$ & $133(1)\cdot 10$ & 69.7(1)\\        
	{\tt WHIZARD}             & $106(1)\cdot10^{2}$   & $133(1)\cdot 10$ & 69.7(2)\\
	\hline 
\end{tabular} 
\end{table}

\begin{table}[ht!]
\centering
\caption{
The total cross section of process (\ref{ccA}) in picobarns at a center-of-mass energy of 24 GeV for different combinations of helicities $\chi_1$ and $\chi_2$.}
\label{hel24}
    \begin{tabular}{lllll}
    \hline\hline
    $\chi_{1}\chi_{2}$ & $+ +$ & $- -$ & $+ -$ & $- +$ \\
    \hline
    {\tt ReneSANCe}	          & $167(1)\cdot 10$ & $167(1)\cdot 10$ & $197(1)\cdot 10^{2}$ & $197(1)\cdot 10^{2}$ \\  
    {\tt CalcHEP}             & $167(1)\cdot 10$ & $167(1)\cdot 10$ & $197(1)\cdot 10^{2}$ & $197(1)\cdot 10^{2}$\\        
	{\tt WHIZARD}             & $166(1)\cdot 10$ & $166(1)\cdot 10$ & $196(1)\cdot 10^{2}$ & $196(1)\cdot 10^{2}$\\
	\hline 
\end{tabular} 
\end{table}
\begin{table}[ht!]
\centering
\caption{
The total cross section of process (\ref{ccA}) in picobarns at a center-of-mass energy of 91 GeV for different combinations of helicities $\chi_1$ and $\chi_2$.}
\label{hel91}
    \begin{tabular}{lllll}
    \hline\hline
    $\chi_{1}\chi_{2}$ & $+ +$ & $- -$ & $+ -$ & $- +$ \\
    \hline
    {\tt ReneSANCe}	          & 15.22(1) &  15.22(1) & $265(1)\cdot 10$  & $265(1)\cdot 10$ \\  
    {\tt CalcHEP}             & 15.21(1)   & 15.21(1) & $266(1)\cdot 10$ & $266(1)\cdot 10$\\        
	{\tt WHIZARD}             & 15.21(1)   & 15.21(1) & $264(1)\cdot 10$ & $264(1)\cdot 10$\\
	\hline 
\end{tabular} 
\end{table}
\begin{table}[ht!]
 \centering
\caption{
The total cross section of process (\ref{ccA}) in picobarns at a center-of-mass energy of 500 GeV for different combinations of helicities $\chi_1$ and $\chi_2$.}
\label{hel500}
    \begin{tabular}{lllll}
    \hline\hline
    $\chi_{1}\chi_{2}$ & $+ +$ & $- -$ & $+ -$ & $- +$ \\
    \hline
    {\tt ReneSANCe}	          & 0.0259(1) & 0.0259(1) & 139(1) & 139(1) \\  
    {\tt CalcHEP}             & 0.0259(1) & 0.0259(1) & 139(1)  & 139(1)\\        
	{\tt WHIZARD}             & 0.0259(1) & 0.0259(2) & 139(1) & 139(1)\\
	\hline 
\end{tabular} 
\end{table}

\begin{figure}[htbp]
    \centering
    \begin{minipage}[t]{0.48\textwidth}
        \centering
        \includegraphics[width=\linewidth]{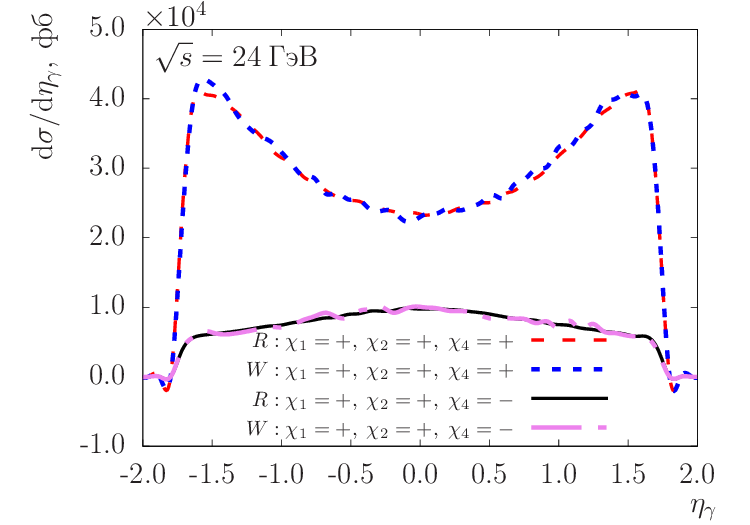}
        \caption{Differential cross section distributions of process (\ref{ccA}) in femtobarns as a function of the photon pseudorapidity, $\eta_{\gamma}$, at a center-of-mass energy of 24 GeV for the helicity combination $\chi_{1} = +$, $\chi_{2} = +$, and photon helicity $\chi_{4} = \pm$. $R$ denotes \texttt{ReneSANCe} and $W$ denotes \texttt{WHIZARD}.}
        \label{fig:1}
    \end{minipage}\hfill
    \begin{minipage}[t]{0.48\textwidth}
        \centering
        \includegraphics[width=\linewidth]{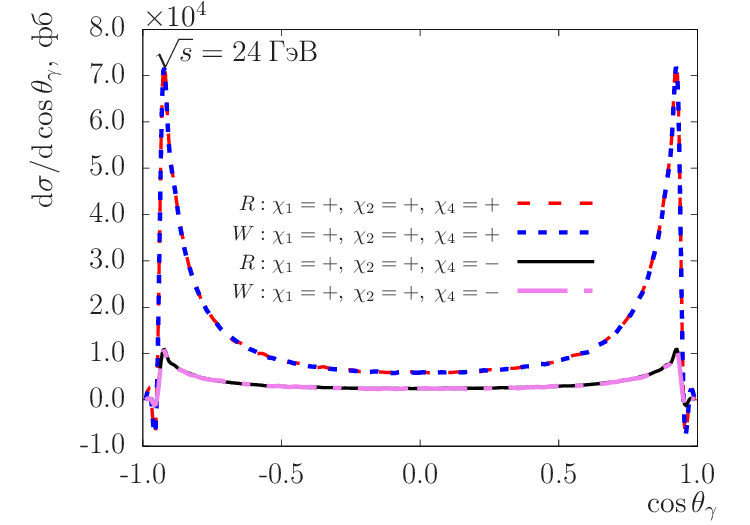}
        \caption{Differential cross section distributions of process (\ref{ccA}) in femtobarns as a function of the photon scattering angle cosine, $\cos\theta_{\gamma}$, at a center-of-mass energy of 24 GeV for the helicity combination $\chi_{1} = +$, $\chi_{2} = +$, and photon helicity $\chi_{4} = \pm$. $R$ denotes \texttt{ReneSANCe} and $W$ denotes \texttt{WHIZARD}.}
        \label{fig:2}
    \vspace{1.5cm}
    \end{minipage}
    \begin{minipage}[t]{0.48\textwidth}
        \centering
        \includegraphics[width=\linewidth]{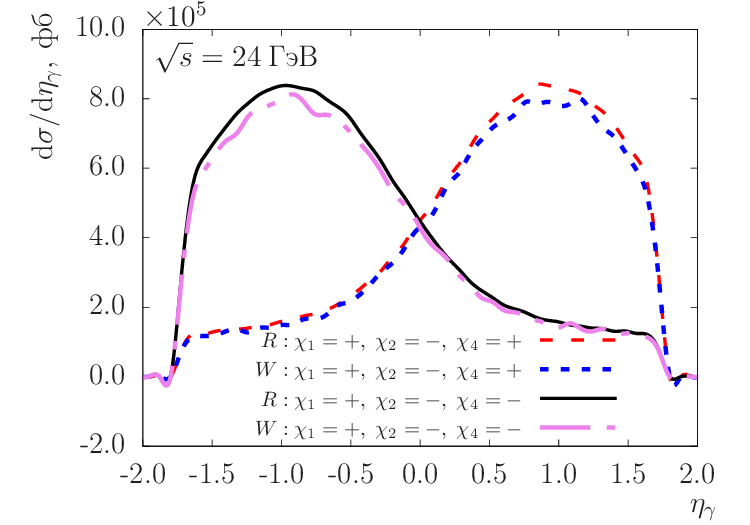}
        \caption{Differential cross section distributions of process (\ref{ccA}) in femtobarns as a function of the photon pseudorapidity, $\eta_{\gamma}$, at a center-of-mass energy of 24 GeV for the helicity combination $\chi_{1} = +$, $\chi_{2} = -$, and photon helicity $\chi_{4} = \pm$. $R$ denotes \texttt{ReneSANCe} and $W$ denotes \texttt{WHIZARD}.}
        \label{fig:3}
    \end{minipage}\hfill
    \begin{minipage}[t]{0.48\textwidth}
        \centering
        \includegraphics[width=\linewidth]{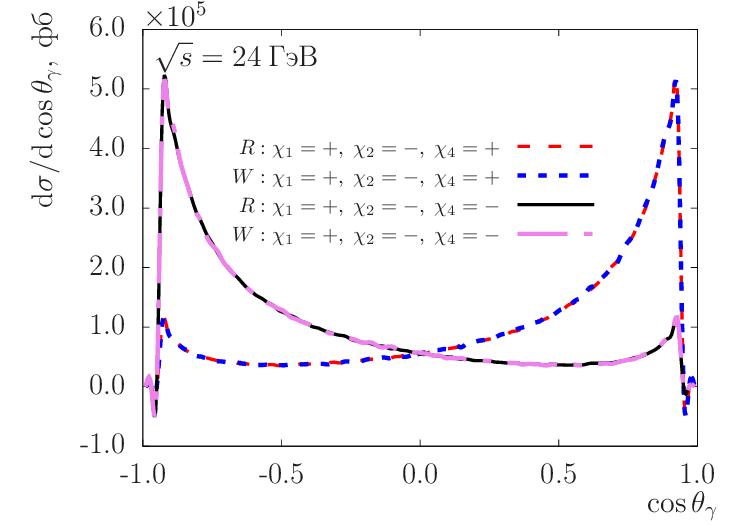}
        \caption{Differential cross section distributions of process (\ref{ccA}) in femtobarns as a function of the photon scattering angle cosine, $\cos\theta_{\gamma}$, at a center-of-mass energy of 24 GeV for the helicity combination $\chi_{1} = +$, $\chi_{2} = -$, and photon helicity $\chi_{4} = \pm$. $R$ denotes \texttt{ReneSANCe} and $W$ denotes \texttt{WHIZARD}.}
        \label{fig:4}
    \end{minipage}
\end{figure}

\begin{figure}[htbp]
    \centering
    \begin{minipage}[t]{0.48\textwidth}
        \centering
        \includegraphics[width=\linewidth]{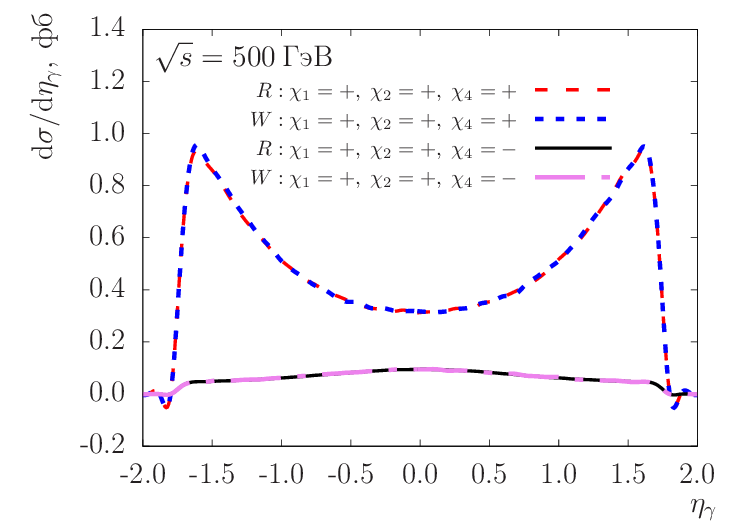}
        \caption{Differential cross section distributions of process (\ref{ccA}) in femtobarns as a function of the photon pseudorapidity, $\eta_{\gamma}$, at a center-of-mass energy of 500 GeV for the helicity combination $\chi_{1} = +$, $\chi_{2} = +$, and photon helicity $\chi_{4} = \pm$. $R$ denotes \texttt{ReneSANCe} and $W$ denotes \texttt{WHIZARD}.}
        \label{fig:5}
    \end{minipage}\hfill
    \begin{minipage}[t]{0.48\textwidth}
        \centering
        \includegraphics[width=\linewidth]{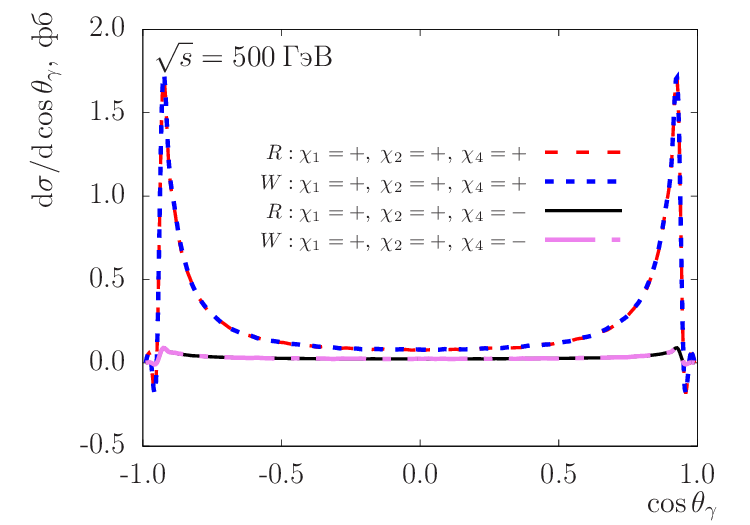}
        \caption{Differential cross section distributions of process (\ref{ccA}) in femtobarns as a function of the photon scattering angle cosine, $\cos\theta_{\gamma}$, at a center-of-mass energy of 500 GeV for the helicity combination $\chi_{1} = +$, $\chi_{2} = +$, and photon helicity $\chi_{4} = \pm$. $R$ denotes \texttt{ReneSANCe} and $W$ denotes \texttt{WHIZARD}.}
        \label{fig:6}
    \vspace{1.5cm}
    \end{minipage}
    \begin{minipage}[t]{0.48\textwidth}
        \centering
        \includegraphics[width=\linewidth]{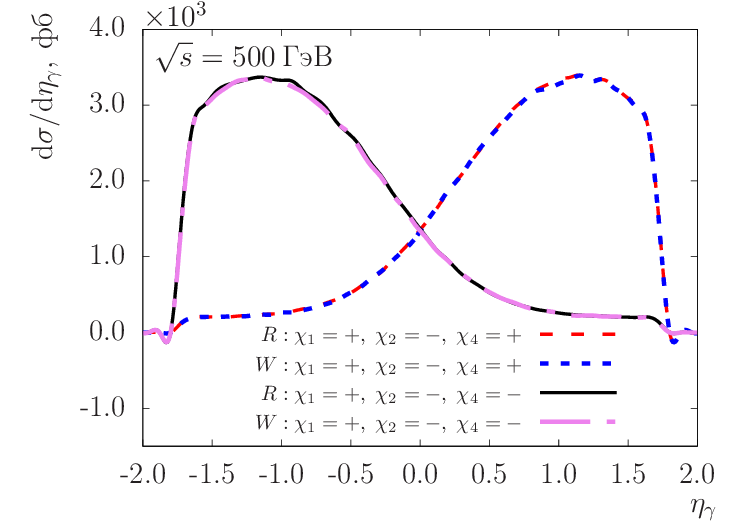}
        \caption{Differential cross section distributions of process (\ref{ccA}) in femtobarns as a function of the photon pseudorapidity, $\eta_{\gamma}$, at a center-of-mass energy of 500 GeV for the helicity combination $\chi_{1} = +$, $\chi_{2} = -$, and photon helicity $\chi_{4} = \pm$. $R$ denotes \texttt{ReneSANCe} and $W$ denotes \texttt{WHIZARD}.}
        \label{fig:7}
    \end{minipage}\hfill
    \begin{minipage}[t]{0.48\textwidth}
        \centering 
        \includegraphics[width=\linewidth]{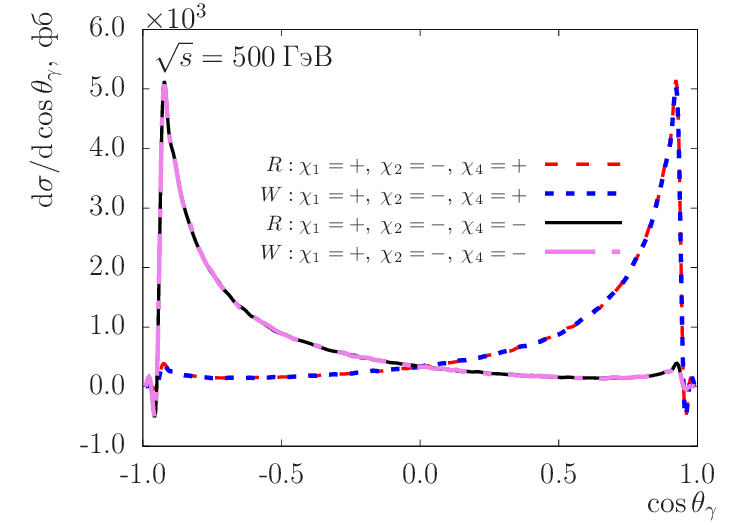}
        \caption{Differential cross section distributions of process (\ref{ccA}) in femtobarns as a function of the photon scattering angle cosine, $\cos\theta_{\gamma}$, at a center-of-mass energy of 500 GeV for the helicity combination $\chi_{1} = +$, $\chi_{2} = -$, and photon helicity $\chi_{4} = \pm$. $R$ denotes \texttt{ReneSANCe} and $W$ denotes \texttt{WHIZARD}.}
        \label{fig:8}
    \end{minipage}
\end{figure}

A comparison was also made of the differential cross section distributions with respect to $\cos\theta_{\gamma}$ and photon pseudorapidity $$\eta_{\gamma}=-\ln\left(\tan \dfrac{\vartheta_{\gamma}}{2}\right)$$ with taking into account the helicities of the photon $\gamma(\chi_{4})$, and $\bar{c}(\chi_{1})$- and $c(\chi_{2})$-quarks at energies of 24 GeV (Figs. \ref{fig:1}, \ref{fig:2}, \ref{fig:3}, \ref{fig:4}) and 500 GeV (Figs. \ref{fig:5}, \ref{fig:6}, \ref{fig:7}, \ref{fig:8}). Good agreement has been achieved between \texttt{ReneSANCe} and \texttt{WHIZARD} for all considered cases. As can be seen, at 24 GeV, the largest contribution corresponds to the helicity combination $\chi_{1} = +$, $\chi_{2} = -$, and $\chi_{4} = \pm$, which exceeds the contribution of the $\chi_{1} = +$, $\chi_{2} = +$, and $\chi_{4} = \pm$ combination by one order of magnitude.
At 500 GeV, a similar pattern is observed, but the difference between the contributions of the aforementioned combinations increases to three orders of magnitude.

\section{Conclusion}
\label{concl}

The primary priority in the development of {\tt SANC} as a tool for the theoretical support of high-energy physics is the creation of Monte Carlo tools incorporating one-loop level corrections, as well as the development of standalone modules for one-loop cross-section components aimed at integration into existing external Monte Carlo event generators.

This article provides a detailed description of the analytical calculation of the amplitude for one of the contributions in one-loop calculations for prompt photon production processes (\ref{qqgA}) within the {\tt SANC} framework. For these processes, the helicity amplitudes of hard bremsstrahlung gluon and photon radiation have been obtained for the first time in the spinor formalism, with taking into account the helicities and masses of all particles.

Numerical calculations and verification have been performed using the example of the quark-antiquark annihilation channel $\bar{c}c$ for various combinations of helicity states and center-of-mass energies. Good agreement has been achieved with the results from \texttt{WHIZARD} and \texttt{CalcHEP}, both for cases considering particle helicity states and for those averaged over them.

The study of prompt photon production processes with allowance for one-loop radiative corrections and helicities of initial and final states is an important part of the ongoing physical analysis at the \texttt{NICA} collider. The results will be used in an extended analysis encompassing all one-loop level contributions to the prompt photon production process in the annihilation channel, and the developed computational modules will be integrated into the next version of the \texttt{ReneSANCe} generator.

\textbf{Acknowledgments.} The author expresses gratitude to the {\tt SANC} project team for fruitful discussions and support during the preparation of this publication.

\end{document}